\documentclass[12pt]{article}
\usepackage{epsfig}
\usepackage{amssymb}
\usepackage{pslatex}

\makeatletter

\@addtoreset{equation}{section}
\makeatother
\newcommand{\bea}{\begin{eqnarray}}
\newcommand{\eea}{\end{eqnarray}}
\newcommand{\be}{\begin{eqnarray}}
\newcommand{\ee}{\end{eqnarray}}

\def\tr{\mathop{\rm Tr}}

\def\fR{{\mathfrak R}}

\def\rmd{{\rm d}}

\begin{document}

\begin{titlepage}
\vskip-0cm
\begin{flushright}
SNUST 080801
\end{flushright}
\centerline{\Large \bf  Integrable Spin Chain}
\vskip0.25cm
\centerline{\Large \bf of}
 \vskip0.25cm
 \centerline{\Large \bf Superconformal U$(M)\times \overline{{\rm U}(N)}$ Chern-Simons Theory }
\vspace{0.75cm}
\centerline{\large Dongsu Bak$^a$, \,\,\, Dongmin Gang$^b$, \,\,\, Soo-Jong Rey$^b$}
\vspace{0.75cm}
\centerline{\sl ${}^a$ Physics Department, University of Seoul, Seoul 130-743 {\rm KOREA}}
\vskip0.25cm
\centerline{\sl ${}^b$ School of Physics \& Astronomy, Seoul National University, Seoul 151-747 {\rm KOREA}}
\vskip0.25cm
\centerline{\tt dsbak@uos.ac.kr \,\,\, arima275@snu.ac.kr \,\,\, sjrey@snu.ac.kr}
\vspace{1.0cm}
\centerline{ABSTRACT}
\vspace{0.75cm}
\noindent
${\cal N}=6$ superconformal Chern-Simons theory with gauge group U$(M)\times \overline{{\rm U}(N)}$ is dual to $N$ M2-branes and $(M-N)$ fractional M2-branes, equivalently, discrete 3-form holonomy at $\mathbb{C}^4/\mathbb{Z}_k$ orbifold singularity. We show that, much like its regular counterpart of $M=N$, the theory at planar limit have integrability structure in the conformal dimension spectrum of single trace operators. We first revisit the Yang-Baxter equation for a spin chain system associated with the single trace operators. We show that the integrability by itself does not preclude parity symmetry breaking. We construct two-parameter family of parity non-invariant, alternating spin chain Hamiltonian involving three-site interactions between ${\bf 4}$ and $\overline{\bf 4}$ of SU(4)$_R$. At weak `t Hooft coupling, we study the Chern-Simons theory perturbatively and calculate anomalous dimension of single trace operators up to two loops. The computation is essentially parallel to the regular case $M=N$. We find that resulting spin chain Hamiltonian matches with the Hamiltonian derived from Yang-Baxter equation, but to the one preserving parity symmetry. We give several intuitive explanations why the parity symmetry breaking is not detected in the Chern-Simons spin chain Hamiltonian at perturbative level. We suggest that open spin chain, associated with open string excitations on giant gravitons or dibaryons, can detect discrete flat holonomy and hence parity symmetry breaking through boundary field.

\end{titlepage}

\section{Introduction}
In continuation of remarkable development by Aharony, Bergman, Jafferis and Maldacena (ABJM)~\cite{Aharony:2008ug}, Aharony, Bergman and Jafferis (ABJ)~\cite{Aharony:2008gk} identified
further examples of AdS/CFT correspondences: three-dimensional ${\cal N}=6$ superconformal Chern-Simons theory with gauge group U(M)$_k\times\overline{\rm U(N)}_{-k}$, where $k$ denotes the Chern-Simons level, is dual to Type IIA string theory on AdS$_4 \times \mathbb{CP}^3$ \cite{Nilsson:1984bj} with $B_{\rm NS}$ holonomy turned on over $\mathbb{CP}_1 \subset \mathbb{CP}_3$.
For consistency with flux quantization of Ramond-Ramond field strengths, the $B_{\rm NS}$ holonomy is not arbitrary but takes a value in $\mathbb{Z}_k$ (measured in string unit). From M-theory viewpoint, the gravity dual background descends from AdS$_4 \times \mathbb{S}^7/\mathbb{Z}_k$ once M-theoretic discrete torsion is turned on over a torsion 3-cycle in
$\mathbb{S}^7/\mathbb{Z}_k$. The corresponding torsion flux takes a value in $H^4 (\mathbb{S}^7/\mathbb{Z}_k, \mathbb{Z}) = \mathbb{Z}_k$. In the limit $M \rightarrow N$, these discrete fluxes are turned off and the new correspondence \cite{Aharony:2008gk} is reduced to the correspondence identified earlier \cite{Aharony:2008ug}~\footnote{In~\cite{Aharony:2008gk}, the authors also proposed AdS/CFT correspondence for orientifold variants with ${\cal N}=5$ superconformal symmetry. In what follows, for concreteness, we shall focus on the subsets with ${\cal N}=6$ superconformal symmetry. Lagrangian of these superconformal field theories were previously studied in~\cite{Hosomichi:2008jb}.}.

The purpose of this paper is to show that, much the same as the ABJM theory~\cite{Minahan:2008hf, Rey1}, the ABJ theory also exhibits integrability structure in the spectrum of anomalous dimensions for single trace local operator~\footnote{For other important works on integrability structure at the weak coupling limit, see~\cite{Nishioka:2008gz, Gaiotto:2008cg}.}.
By extending the computations of \cite{Rey1}, we shall find that the spin chain Hamiltonian that governs two-loop operator mixing and anomalous dimensions in ABJ theory is essentially the same as that of ABJM theory modulo suitable change of perturbative coupling parameters.

We organized this paper as follows. In section 2, in comparison with the ABJM theory, we list several new features of the ABJ theory that will be directly relevant for the quest of integrability structure. In
section 3, we revisit the derivation of integrable spin chain from Yang-Baxter equations. We emphasize that parity symmetry of the spin chain is broken in general. We construct the most general parity non-invariant, integrable spin chain Hamiltonian and show that, up to overall scaling, there are two-parameter family of Hamiltonian.
In section 4, we compute operator mixing and anomalous dimensions of single trace operators at two loops. We find that the resulting Chern-Simons spin chain Hamiltonian coincides with the spin chain
Hamiltonian derived from Yang-Baxter equation. In fact, the Hamiltonian is exactly the same as the Hamiltonian for ABJM theory \cite{Minahan:2008hf, Rey1}, except that the coupling parameter $N^2$ in the ABJM theory is now replaced by $MN$. In section 5, we offer several arguments why the spin chain Hamiltonian does not detect parity violation effect of the $B_{\rm NS}$ holonomy and illustrate them by studying giant magnon. We also suggest that the discrete holonomy may be visible for an open spin chain associated with open string attached to giant graviton or dibaryon operators.

\section{Aspects of ABJ Theory}

In the ABJ theory, since the number of fractional branes is a new parameter added to the ABJM theory, there are three coupling parameters, $M, N, k$. In contrast to ${\cal N}=4$ super Yang-Mills theory, a unique feature of the ABJ theory (as well as ABJM theory) is that the coupling parameters are all integer-valued. In this section, we elaborate several notable aspects of the ABJ theory that will become relevant for later investigation of integrability. For these, we shall take the generalized `t Hooft planar limit (in the convention $M \ge N$):
\bea
M, \,\,\, N, \,\,\, k \longrightarrow \infty \qquad
\mbox{with} \qquad \lambda \equiv {N \over k}, \quad \overline{\lambda} \equiv {M \over k}, \quad b\equiv {(M-N) \over k} \quad \mbox{fixed} \, ,
\label{thooftparameter}
\eea
though some of the results are extendible beyond this limit. Among these, the parameter $b$ is parity-odd and measures parity symmetry breaking effects in ABJ theory.

In this section, we elaborate several salient features of the ABJ theory that will become
directly relevant for our foregoing investigation on integrability structure.
\begin{list}{$\bullet$}{}
\item
From the viewpoint of M2-branes probing $\mathbb{C}^4/\mathbb{Z}_k$ orbifold singularity, the ABJ theories arise when, in addition to $N$ M2-branes, $(M-N)$ fractional M2-branes are localized at the orbifold singularity. In the much studied situation of $N$ D3-branes probing ${\cal M}_5/\Gamma$ orbifold singularity, adding fractional D-branes~\cite{Gimon:1996rq, Douglas:1996xg} at the orbifold singularity~\cite{Gubser:1998fp, Dasgupta:1999wx} led to running of otherwise constant gauge coupling parameter and hence to loss of the conformal invariance. This implies that, in the large $N$ limit, gravity dual background is deformed away from AdS$_5 \times {\cal M}_5/\Gamma$~\cite{Klebanov:2000nc, Klebanov:2000hb}. In the ABJ theories, even though fractional M2-branes are introduced, the supergravity background is not deformed at all and retains AdS$_4 \times \mathbb{S}^7/\mathbb{Z}_k$. We can understand this curious feature from noting that the gauge-gravity correspondence at hand involves superconformal Chern-Simons theories. In the latter theories, coupling parameters $M, N, k$ are all quantized to integer values. Therefore, at quantum level, these coupling parameters cannot possibly run under renormalization group flow. As such, we expect that operator mixing and anomalous dimensions of gauge invariant composite operators are still organized in the planar limit $M, N \rightarrow \infty$ as an analytic perturbative series expansion of the `t Hooft coupling parameters (\ref{thooftparameter}) within finite radius of convergence~\footnote{Recall that, in ${\cal N}=4$ super Yang-Mills theory, the radius of convergence of planar expansion is $|\lambda| = \pi^2$~\cite{Beisert:2006ez}.}.

\item
Introducing fractional M2-branes or turning on $B_{\rm NS}$ holonomy, the parity symmetry is broken in the supergravity dual background and, in accordance with AdS/CFT correspondence, in the superconformal Chern-Simons theory. Apparently, parity transformation maps the Chern-Simons parameters by $k$ to $-k$ while holding $M, N$ fixed. In the planar limit (\ref{thooftparameter}), this maps $b$ to $-b$ while holding $\lambda, \overline{\lambda}$ fixed. We recall that the parity symmetry $P$ in ABJM theory was defined as, under $x^m \rightarrow - x^m$,
\bea
P: \hskip1cm (A_m, \overline{A}_m, Y^I, Y^\dagger_I) \qquad \rightarrow \qquad ( -\overline{A}_m, -A_m, Y^\dagger_I, Y^I) \label{abjmparity}.
\eea
In particular, since $Y^I \leftrightarrow Y^\dagger_I$, the parity exchanges the two isomorphic gauge groups, U(N) and $\overline{\rm U(N)}$. In the
corresponding spin chain, this was identified with interchange of two interlaced chains of ${\bf 4}$'s and
$\overline{\bf 4}$'s. From the viewpoint of SU(4) symmetry, this is equivalent to charge conjugation. As such, the above $(2+1)$-dimensional parity transformation acts on the spin chain as $(1+1)$-dimensional parity transformation:
 \bea
P: \hskip1cm \mbox{Tr} (Y^{I_1} Y^\dagger_{J_1} \cdots Y^{I_n} Y^\dagger_{J_n}) \qquad
 \rightarrow \qquad \mbox{Tr} (Y^{J_n} Y^\dagger_{I_n} \cdots Y^{J_1} Y^\dagger_{I_1}).
 \label{abjmspinchainparity}\eea
 Hence, under this generalized parity transformation, the ABJM theory and the corresponding spin chain were manifestly invariant. Now, in the ABJ theory, the above parity transformation cannot possibly be a symmetry since, among others, the two gauge groups are different and cannot be exchanged. In fact, as we shall see below, the parity maps one ABJ theory with a given gauge group to another with different gauge group. Therefore, the newly identified correspondences of the ABJ theoy offer an excellent playground for exploring physics associated with parity symmetry and its breaking. In the quest of the integrability, this also raises very interesting issues: Is integrability compatible with parity symmetry breaking? Is parity symmetry breaking always reflected in the associated spin chain system? If so, what kind of spin chain Hamiltonian and higher conserved charges does it lead to? How visible is the parity symmetry breaking effect at weak and strong
`t Hooft coupling regimes?

In the ABJM theory, the parity transformation mapped the theory to itself, viz. parity invariant. In ABJ theory, the parity transformation relates one theory to another in a rich manner. To see this, recall that the ABJ theory with U(M)$_k\times\overline{\rm U(N)}_{-k}$ gauge group is realizable via regular and fractional D3-branes threading between two diametrically separated $(p,q)$-branes of charge $(1,0)\oplus(1,k)$. If we adiabatically move $(1,0)$-brane and $(1,k)$-brane relatively and exchange their locations, the $(M-N)$ fractional D3-branes will disappear on one interval of the two $(p,q)$-branes and the $k-(M-N)$ fractional D3-branes are created on the other interval~\cite{Hanany:1996ie, reyunpub98}. Therefore, the original D3-branes $M_k \oplus N_{-k}$ is transformed to the one $N_k \oplus (N + k - (M-N))_{-k}$. Combining also with the parity transformed theory, we then have equivalence relations:
\bea
{\rm U(M)}_k \times \overline{{\rm U(N)}}_{-k} \,\,\, \simeq \,\,\, {\rm U(N)}_k \times\overline{{\rm U}(2N - M + k)}_{-k} \,\,\, \simeq \,\,\, {\rm U}(N)_k \times \overline{{\rm U}(M)}_{-k}.
\eea
Notice that the relation is entirely among Chern-Simons theories. In particular, the middle theory
is always strongly coupled, since the `t Hooft coupling of the second gauge group is always great than unity. This is exciting; in the quest of integrability and its interpolation between weak and strong coupling limits, the above equivalence relations may provide a new useful trick to extract physical observables such as (generalized) scaling functions not just at weak and strong `t Hooft coupling limits but also at ${\cal O}(1)$ regimes (albeit the drawback that these are all in lower-dimensional field theories).

\item
The number of fractional M2-branes is not arbitrary but is limited to $0 \le (M-N) \le k$. This is most clearly seen from decoupling limit of $(M-N)$ many fractional M2-branes from $N$ many regular M2-branes.
Low-energy dynamics of the fractional brane is described by ${\cal N}=3$ supersymmetric pure Chern-Simons theory with gauge group U$(M-N)_k$. Quantum mechanically, the Chern-Simons level $k$ of this theory ought to remain the same. This can be understood, for example, from the brane construction mentioned above: at all scales of D3-brane dynamics, the $(p,q)$-brane charges are held fixed. But such a non-renormalization property turns out possible only if $(M-N) \le k$. To see this, we can sequentially integrate out superpartners of the gauge fields first and then the gauge field. The first step yields a bosonic pure Chern-Simons theory with gauge group U$(M-N)_{k'}$ where $k' = k - (M-N)\mbox{sign}(k)$. The second step shifts the Chern-Simons level further to $k'' = k' + (M-N) \mbox{sign}(k')$. We see that the Chern-Simons level at quantum level $k''$ remains the same as the classical one $k$ if and only if $(M-N) \le k$. Stated in the planar limit (\ref{thooftparameter}), this quantum consistency restricts the parity-odd coupling parameter $b$ to take values less than unity. In particular, in the strong coupling limit where supergravity dual description is effective, we expect that parity symmetry breaking effect is completely invisible since $b \ll \lambda, \overline{\lambda}$.

\item
AdS/CFT correspondence asserts that gauge invariant, single trace operators in the ABJ theory are dual to free string excitation modes in AdS$_4 \times \mathbb{CP}^3$ with $B_{\rm NS}$ holonomy over $\mathbb{CP}^1$, valid at weak and strong `t Hooft coupling regime, respectively. In particular, conformal dimension of the operators should match with excitation energy of the string modes~\footnote{ Classical integrability of semiclassical string on AdS$_4 \times\mathbb{CP}^3$ was argued in~\cite{Arutyunov:2008if, Stefanski:2008ik, Gromov:2008bz, Rey1}. We find that, even though discrete $B_{\rm NS}$ holonomy is turned on, integrability extends trivially.}.
As summarized in the Appendix, the ABJ theory with gauge group U(M)$\times\overline{\rm U(N)}$ is not much different from the ABJM theory: it possesses ${\cal N}=6$ superconformal symmetry with SO(6)$\simeq$SU(4) R-symmetry and contains two sets of bi-fundamental scalar fields $Y^I, Y^\dagger_I$ $(I=1,2,3,4)$ that transform as ${\bf 4}, \overline{\bf 4}$ under SU(4) and as $({\bf M}, \overline{\bf N})$ and $(\overline{\bf M}, {\bf N})$ under the gauge group U(M)$\times\overline{\rm U(N)}$. Therefore, the single trace operators still take the form:
\bea
{\bf \cal O} &=& \mbox{Tr} (Y^{I_1} Y^\dagger_{J_1} \cdots Y^{I_L} Y^\dagger_{J_L} ) C^{J_1 \cdots J_L}_{I_1 \cdots I_L} \nonumber \\
&=& \overline{\mbox{Tr}} (Y^\dagger_{J_1} Y^{I_1} \cdots Y^\dagger_{J_L} Y^{I_L}) C^{J_1 \cdots J_L}_{I_1 \cdots I_L} \ , \label{STO}
\eea
where now Tr and $\overline{\rm Tr}$ refer to trace over U(M) and $\overline{\rm U(N)}$, respectively. The chiral primary operators, corresponding to the choice of (\ref{STO}) with $C^{J_1 \cdots J_L}_{I_1 \cdots I_L}$ totally symmetric in both sets of indices and traceless, form the lightest states. They correspond to the Kaluza-Klein supergravity modes on gravity dual background. Since the gravity dual of the ABJ theory is still the same as the ABJM theory, viz. AdS$_4 \times \mathbb{CP}^3$, ABJ claims that the spectrum of non-baryonic chiral primary operators is independent of $(M-N)$ and hence $b$~\footnote{ABJ argues that spectrum of baryonic chiral primary operators depends on $b$, so deviates from the ABJM theory. We shall revisit excitation of baryonic operators later in Section 5}. This entails an interesting question: is the spectrum and the spectral distribution of all single trace operators, not just chiral primary operators, independent of $b$?

\end{list}
\section{Integrable Spin Chain from Yang-Baxter}
Given the distinctive features as above, does the ABJ theory also exhibit an integrability structure? If so, since the ABJ theory is parity non-invariant, we must address if integrability structure is compatible with parity symmetry breaking. Paying attention to this, in this section, we revisit derivation of the spin chain Hamiltonian associated with the single trace operators (\ref{STO}).

Operator mixing under renormalization and their evolution in perturbation theory is describable by
a spin chain of total length $2L$. From the structure of operators (\ref{STO}), we see that the prospective spin chain involves two types of SU$_R$(4) spins: ${\bf 4}$ at odd lattice sites and $\overline{\bf 4}$ at even lattice sites. Since we are dealing with gauge invariant operators, these considerations are independent of actual values and relations of $M, N$ in so far as they are taken to the planar limit, $M, N \rightarrow \infty$. It is thus natural to expect that the prospective spin chain is again the same `alternating SU(4) spin chain' of interlaced
${\bf 4}$ and $\overline{ \bf 4}$ as that featured in the ABJM theory~\cite{Minahan:2008hf, Rey1}.

Identification of prospective spin system starts with solving inhomogeneous Yang-Baxter equations of SU$_R$(4) $\mathfrak{R}$-matrices with varying representations on each site. Following the general procedure \cite{deVega:1991rc}, the Yang-Baxter equations were solved in~\cite{Minahan:2008hf, Rey1}. In this section, we shall repeat the procedure of~\cite{Rey1} and emphasize that the putative SU(4) spin chain is the 'alternating spin chain' involving next-to-nearest neighbor interactions
{\sl and} that the integrable spin chain extracted from the Yang-Baxter equations in general breaks the parity symmetry.

As the elementary constituents are in the representations ${\bf 4}, \overline{\bf 4}$ of SU(4)$_R$, we start with R-matrices ${\mathfrak R}^{\bf 44}(u)$ and ${\mathfrak R}^{\bf 4\bar{\bf 4}}(u)$, where the upper indices denote SU(4) representations of two spins involved in `scattering process' and $u, v$ denote spectral parameters. We demand these R-matrices to satisfy two sets of Yang-Baxter equations:
\bea
&& \fR^{\bf 44}_{12}(u-v)\, \fR^{\bf 44}_{13}(u)\, \fR^{\bf 44}_{23}(v)
= \fR^{\bf 44}_{23}(v)\, R^{\bf 44}_{13}(u)\, R^{\bf 44}_{12}(u-v) \label{444YBE} \\
&&
\fR^{\bf 44}_{12}(u-v) \, \fR^{\bf 4\bar{\bf 4}}_{13}(u)\, \fR^{\bf 4\bar{\bf 4}}_{23}(v)
= \fR^{\bf 4\bar{\bf 4}}_{23}(v)\, \fR^{\bf 4\bar{\bf 4}}_{13}(u)\, \fR^{\bf 44}_{12}(u-v)
\label{4bar4bar4YBE} \eea
Here, the lower indices $i,\,j$ denote that the $\fR$ matrix is acting
on $i$-th and $j$-th site $V_i \otimes V_j$ of the full tensor product Hilbert space $V_1 \otimes V_2 \otimes \cdots \otimes V_{2L}$. We find that the R-matrices solving (\ref{444YBE}, \ref{4bar4bar4YBE}) are the well
known ones:
\be
\fR^{\bf 44}(u) = u \mathbb{I} + \mathbb{P} \qquad \mbox{and} \qquad
\fR^{\bf 4\bar{\bf 4}}(u)= -(u+2+\alpha) \mathbb{I} + \mathbb{K}\,
\ee
where $\alpha$ is an arbitrary constant to be determined later. Here, we have introduced identity operator $\mathbb{I}$, trace operator $\mathbb{K}$, and permutation operator $\mathbb{P}$:
\be
(\mathbb{I}_{k \ell})^{I_k I_\ell}_{J_k J_\ell} = \delta^{I_k}_{J_k} \delta^{I_\ell}_{J_\ell} \qquad  \qquad
(\mathbb{K}_{k \ell})^{I_k I_\ell}_{J_k J_\ell} = \delta^{I_k I_\ell}
\delta_{J_k J_\ell } \qquad \qquad
(\mathbb{P}_{k \ell})^{I_k I_\ell}_{J_k J_\ell} = \delta^{I_k}_{J_\ell} \delta^{I_\ell}_{J_k}
\ ,
\label{braiding}
\ee
acting as braiding operations mapping tensor product vector space $V_k \otimes V_\ell$ to itself.

Similarly, we also construct another set of R-matrices $\fR^{\bar{\bf 4}\bar{\bf 4}}(u)$ and $\fR^{\bar{\bf 4}\bf 4}(u)$ for 'scattering process' of the specified quantum number constituents. They will generate another alternative spin chain system. Demanding them to fulfill the respective
Yang-Baxter equations:
\bea
&& \fR^{\bar{\bf 4}\bar{\bf 4}}_{12}(u-v)\, \fR^{\bar{\bf 4}\bar{\bf 4}}_{13}(u)
\, \fR^{\bar{\bf 4}\bar{\bf 4}}_{23}(v)
= \fR^{\bar{\bf 4}\bar{\bf 4}}_{23}(v)\, \fR^{\bar{\bf 4}\bar{\bf 4}}_{13}(u)\,
\fR^{\bar{\bf 4}\bar{\bf 4}}_{12}(u-v) \label{bar444YBE} \\
&&
\fR^{\bf 44}_{12}(u-v)\, \fR^{\bar{\bf 4}\bf 4}_{13}(u)\, \fR^{\bar{\bf 4}\bf 4}_{23}(v)
= \fR^{\bar{\bf 4}\bf 4}_{23}(v) \, \fR^{\bar{\bf 4}\bf 4}_{13}(u)\,  \fR^{\bf 44}_{12}(u-v)
\label{bar4bar4YBE}
\eea
we find that the solution is given by
\be
\fR^{\bar{\bf 4}\bar{\bf 4}}(u) = u \mathbb{I} + \mathbb{P} \qquad \mbox{and} \qquad
\fR^{\bar{\bf 4}\bf 4}(u)= -(u+2+ \bar{\alpha})\mathbb{I} + \mathbb{K}\,,
\ee
where 
$\bar{\alpha}$ is an arbitrary constant.

In the two sets of Yang-Baxter equations, the constants $\alpha, \bar{\alpha}$ are undetermined. We shall now restrict them by requiring unitarity. The unitarity of the combined spin chain system sets the following conditions:
\bea
&& \fR^{\bf 44}(u)\, \fR^{\bf 44}(-u)\  = \rho(u)  \mathbb{I}\, \ \ \nonumber\\
&& \fR^{\bar{\bf 4}\bar{\bf 4}}(u)\, \fR^{\bar{\bf 4}\bar{\bf 4}}(-u)\ = \bar{\rho}(u) \ \mathbb{I} \nonumber\\
&& \fR^{\bf 4\bar{\bf 4}}(u)\, \fR^{\bar{\bf 4}\bf 4}(-u)\  = \sigma(u) \  \mathbb{I}
\eea
where $\rho(u) = \rho(-u), \bar{\rho}(u) = \bar{\rho}(-u), \sigma(u)$ are  $c$-number functions.
It follows that the first two unitarity conditions are indeed satisfied for any $\alpha, \bar{\alpha}$, while the last unitarity condition is is satisfied only if $\alpha= -\bar{\alpha}$. Without loss of generality, we shall set $\alpha=- \bar{\alpha}=0$.

Viewing (\ref{STO}) as $2L$ sites of alternating ${\bf 4}$ and $\bar{\bf 4}$ in a row,
we 
introduce monodromy T-matrix
%
%
\be
T_0 (u,a)= \fR^{\bf 44}_{01}(u) \fR^{\bf 4{\bar{\bf 4}}}_{02}(u+a)
\fR^{\bf 44}_{03}(u) \fR^{\bf 4{\bar{\bf 4}}}_{04}(u+a)\cdots
\fR^{\bf 44}_{02L-1}(u) \fR^{\bf 4{\bar{\bf 4}}}_{02L}(u+a)\,,
\ee
for one alternating chain and another monodromy T-matrix
\be
\overline{T}_0 (u,\bar{a})
= \fR^{\bar{\bf 4}\bf 4}_{01}(u+\bar{a})
 \fR^{\bar{\bf 4}\bar{\bf 4}}_{02}(u)
 \fR^{\bar{\bf 4}\bf 4}_{03}(u+\bar{a})
\fR^{\bar{\bf 4}{\bar{\bf 4}}}_{03}(u)\cdots
\fR^{\bar{\bf 4}\bf 4}_{02L-1}(u+\bar{a})
\fR^{{\bar{\bf 4}}\bar{\bf 4}}_{02L}(u) \,,
\ee
for the other alternating chain. The spectral parameters $a, \bar{a}$ are a priori independent since the two spin chains are independent. Yet, intuitively, we expect they are related each other since every lattice site of the inhomogeneous spin chain must have a unique spectral parameter. For now, we shall proceed without a priori such an input and verify that the two are indeed related as an outcome of derivation of the Hamiltonian. Both monodromy T-matrices are defined with respect to an  auxiliary zeroth space. These monodromy T-matrices can be shown to fulfill the Yang-Baxter equations:
\be
\fR^{\bf 44}_{00'}(u-v) T_0 (u,a) T_{0'} (v,a)=
 T_{0'} (v,a)  T_{0} (u,a) \fR^{\bf 44}_{00'}(u-v)\,,
\ee
\be
\fR^{\bar{\bf 4}\bar{\bf 4}}_{00'}(u-v) \overline{T}_0 (u,\bar{a})
\overline{T}_{0'} (v,\bar{a})=
 \overline{T}_{0'} (v,\bar{a})  \overline{T}_{0} (u,\bar{a})
\fR^{\bar{\bf 4}\bar{\bf 4}}_{00'}(u-v)\,.
\ee
and
\be
\fR^{{\bf 4}\bar{\bf 4}}_{00'}(u-v+a) {T}_0 (u, {a})
\overline{T}_{0'} (v,-a)=
 \overline{T}_{0'} (v,-a)  {T}_{0} (u,{a})
\fR^{{\bf 4}\bar{\bf 4}}_{00'}(u-v+a)\,.
\ee

We also define transfer matrix by taking trace of the T matrix over the auxiliary space:
\be
\tau^{\rm alt}(u,a)=\tr_0 {T}_0 (u, {a})\,.
\ee
and
\be
\overline{\tau}^{\rm alt}(u,\bar{a})
= \tr_0\, \overline{T}_0 (u, \bar{a}) \, .
\ee
It then follows from the Yang-Baxter equations that
\bea
&&
[\tau^{\rm alt}(u,a),\tau^{\rm alt}(v,a)]=0
\nonumber\\
&&
[\overline\tau^{\rm alt}(u,\bar{a}),\overline\tau^{\rm alt}(v, \bar{a})]=0\,,
\eea
and
\bea
\hskip1.3cm
[\tau^{\rm alt}(u,a),\bar\tau^{\rm alt}(v, -{a})]=0 \,.
\eea
Here, in the first two equations, $a, \bar{a}$ are arbitrary and
denote two undetermined spectral parameters.
These parameters are restricted further if we demand the last equation
to hold. We showed in~\cite{Rey1} that the two alternating transfer matrices
commute each other if and only if  $ \bar{a} = -a$.

Commuting set of conserved charges are obtained~~\footnote{The following derivation of Hamiltonian is valid only for $L \ge 2$. This means that the energy eigenvalues of the following Hamiltonian
for the case $L=1$ do not agree with true energy eigenvalues.} from moments of the transfer matrix with respect to the spectral parameter $u$. By definition, the Hamiltonian is obtained from the first moment of $\tau^{\rm alt}$: $H \equiv \rmd \log \tau^{\rm alt}(u, a)|_{u=0}$ where $\rmd \equiv \partial / \partial u$. The computational procedure is standard in the context of alternating spin chain and straightforward. After some computation, we found the ${\bf 4}\overline{\bf 4}$ spin chain Hamiltonian acting on ${\bf 4}$ residing sites as
\bea
H^{\rm alt}(a) = \sum_{\ell=1}^L H_{2 \ell - 1}(a) \nonumber
\eea
where
\bea
H_{2\ell - 1}(a)
&=& -(2-a) \mathbb{I} -(4-a^2) \mathbb{P}_{2\ell-1,  2\ell+1} \nonumber \\
&- & (a-2)  \mathbb{P}_{2\ell-1,  2\ell+1}
\mathbb{K}_{2\ell-1,  2\ell}
+(a+2)    \mathbb{P}_{2\ell-1,  2\ell+1}
\mathbb{K}_{2\ell,  2\ell+1} \, ,
\eea
Here, we scaled the Hamiltonian by multiplying $(a^2-4)$. By the same procedure, from the first moment of $\overline{\tau}^{\rm alt}$: $H =\equiv \rmd \log \overline{\tau}^{\rm alt}(v, \bar{a})|_{v=0}$, we also found the Hamiltonian for the $\overline{\bf4}{\bf 4}$ spin chain acting on $\overline{\bf 4}$ sites as
\bea
\overline{H}^{\rm alt}(a) = \sum_{\ell = 1}^L H_{2 \ell}(a) \nonumber
\eea
where
\bea
\overline{H}_{2 \ell}(a)
&=& -(2+a) \mathbb{I} -(4-a^2) \mathbb{P}_{2\ell,  2\ell+2} \nonumber \\
&+& (a+2)  \mathbb{P}_{2\ell,  2\ell+2}
\mathbb{K}_{2\ell,  2\ell+1}- (a-2)  \mathbb{P}_{2\ell,  2\ell+2}
\mathbb{K}_{2\ell+1,  2\ell+2}
\, ,
\eea
where we have replaced $\bar{a}$ by $a$ using
the relation $\bar{a}=-a$. See \cite{Rey1} for details of the derivation.

To have the spin chain Hamiltonian hermitian, as shown in~\cite{Rey1}, we choose the parameter $a$ purely imaginary, $a = i \gamma$. Moreover, since there is no parity symmetry mapping even sites
to odd sites or vice versa, we can have different coupling parameters and different ground state
energy density. Thus, the most general integrable spin chain Hamiltonian reads
\bea
H_{\rm YBE} = \sum_{\ell = 1}^{L} \Big[ J_o \left( H_{2 \ell-1}(\gamma) - \epsilon_o \mathbb{I} \right)
+  J_e \left( \overline{H}_{2 \ell}(\gamma) - \epsilon_e \mathbb{I}\right) \Big].
\label{HfromYBE}
\eea
Here, having two mutually commuting spin chain Hamiltonian by our choice of the spectral parameters, we introduced two coupling parameters $J_e, J_o$ and two ground-state energy parameters $\epsilon_e, \epsilon_o$ for the even and the odd alternate spin chains, respectively. Overall, the spin chain Hamiltonian depends on five parameters $(J_e, \epsilon_e), (J_o, \epsilon_o)$ and $\gamma$. Some of these parameters can be fixed from considerations of
underlying physics of the system. Overall scale can be set to a choice of convention. If we invoke
supersymmetry, the ground-state energy parameters can be fixed by demanding that all chiral primary operators have vanishing energy. This still leaves out two free parameters in the Hamiltonian. For
general choice of these two parameters, the integrable spin chain Hamiltonian (\ref{HfromYBE}) is
parity non-invariant.

In the ABJM theory of $N=M$, the single trace operators had the exchange symmetry ${\bf 4} \leftrightarrow \overline{\bf 4}$. This is the same as the charge conjugation symmetry which entered in the definition of the parity transformation as given in (\ref{abjmparity}) and (\ref{abjmspinchainparity}). We thus put $a=i 0$ in that case. Here, however, since the parity symmetry is broken in the ABJ theory, a priori, there is no reason we stick to this case. This implies that the spin chain Hamiltonian for the ABJ theory may belong to a family of Hamiltonian of the above type. In particular, generically, the parity symmetry is broken.

Despite all these, in the next section, we shall find that the Hamiltonian that actually arise from the planar perturbation theory turns out:
\bea
H_{\rm CS} = {1 \over 4} \lambda \overline{\lambda} \sum_{\ell=1}^{2L} H_{\ell, \ell+1, \ell+2}
\eea
with
\bea
H_{\ell, \ell+1, \ell+2}= \Big[ 4\mathbb{I} - 4 \mathbb{P}_{\ell , \ell+2} + 2  \mathbb{P}_{\ell , \ell+2}
\mathbb{K}_{\ell , \ell+1} + 2 \mathbb{P}_{\ell,  \ell+2}
\mathbb{K}_{\ell+1,  \ell+2} \Big] \
\label{YBEhamiltonian}
\eea
viz. the choice $J_e = J_o = \lambda \overline{\lambda}/4$, $\epsilon_e = \epsilon_o = -6$ and $\gamma = 0$ in (\ref{HfromYBE}). This Hamiltonian
is exactly the same as the spin chain Hamiltonian of ABJM theory except that the coupling parameter
$\lambda^2$ is replaced by $\lambda \overline{\lambda}$. In particular, the Hamiltonian is
completely parity invariant. Stated otherwise, the parity non-invariance of the ABJ theory is not reflected in the spin chain Hamiltonian associated with the single trace operators. We shall discuss reasons behind this in section 5.

\section{Integrable Spin Chain from Chern-Simons}

In this section, we describe the
two loop spin chain Hamiltonian by the direct evaluation of
the anomalous dimension matrix of the suggested
operators.

In general, as well understood from general considerations
of the renormalization theory, the divergence in one-particle
irreducible diagrams with one insertion of a composite operator
contain divergences that are proportional to other
composite operators. Therefore, at each order in
perturbation theory, all composite operators
must be renormalized simultaneously. In addition,
the wave function renormalization of elementary
fields needs to be taken into account. This leads to the
general structure of the renormalization
matrix:
\bea
{\cal O}^A_{\rm bare} (Y_{\rm bare}, Y^\dagger_{\rm bare}) =
\sum_B {Z^A}_B
{\cal O}^B_{\rm ren} (Z Y_{\rm ren}, Z Y^\dagger_{\rm ren})
\eea
For the operators we are interested in,
this takes the form of
\bea
{\cal O}^A_{\rm bare} = \sum_B {Z^A}_B (\Lambda
) {\cal O}^B_{\rm ren}
\eea
with the UV cut-off scale $\Lambda$.
Therefore,
the anomalous dimension matrix $\Delta$ is given by
\bea
\Delta = {{\rm d} \log Z \over {\rm d} \log \Lambda}.
\eea
Below we shall compute anomalous dimension matrix of
the following single trace operator (\ref{STO}) in the basis:
\bea
{\cal O}^{(I)}_{(J)} = {\rm Tr}\, \Big(
Y^{I_1} {Y}^\dagger_{ J_1 } Y^{I_2} {Y}^\dagger_{J_2}\cdots
Y^{I_L} {Y}^\dagger_{J_L} \Big) \ .
\eea
The action for the ${\cal N}=6$ U(M)$\times \overline{\rm U(N)}$
superconformal Chern-Simons theory is the same as that of the
ABJM theory except the change in the gauge symmetry and the matter
representation. We relegate its detailed structure to the appendix.

Basically, the Feynman diagrams and integrals
for the U(M)$\times\overline{\rm U(N)}$
theory with $M\neq N$ (ABJ theory) are not much
different from those of the $M=N$ one (ABJM theory). The general power counting
argument shows that the logarithmic divergence arises only at even loops.
Therefore, nontrivial contribution to the anomalous
dimension again starts at two-loop order. For the bi-fundamental
matter fields with indices $(m,\,\, \bar{n})$,
any loops of the Feynman diagram
involve a sum over the fundamental index $m$ or the anti-fundamental
index $\bar{n}$ giving the factor $M$ and $N$ respectively.
Then the planar diagrams are now organized as a double
power series of the two `t Hooft parameters
$\lambda$ and $\bar\lambda$.
At two loops, the general planar contributions include
terms proportional to
$\lambda^2$, ${\bar\lambda}^2$ and
$\lambda\bar\lambda$.
As we shall explain below, for two-loop anomalous dimension matrix,
we find that all the purely fundamental (${\bar\lambda}^2$)
and purely anti-fundamental ( ${\lambda}^2$)
contributions cancel among themselves and the remaining mixed contributions
lead to the two-loop Hamiltonian:
\be
H_{\rm 2-loops}  = {\lambda\bar{\lambda} }
\sum_{\ell = 1}^{2L} \Big[\mathbb{I} - \mathbb{P}_{\ell, \ell+2} +
{1 \over 2} \mathbb{P}_{\ell, \ell+2} \mathbb{K}_{\ell,\ell+1} +
{1 \over 2} \mathbb{P}_{\ell, \ell+2} \mathbb{K}_{\ell+1, \ell+2} \Big]
\label{twoloops}
\ee
which is integrable clearly.

In this section, we explain derivation of the above Hamiltonian, not by
repeating all the computation and but just counting $\lambda$ and $\bar\lambda$
factors based on the computation of Ref.~\cite{Rey1}. Except
these extra counting factors, all the remaining Feynman
integrals are found to have the same expressions. In particular, there is
no extra diagram that contribute to the Hamiltonian when M is taken different
from N.

\begin{figure}
\begin{center}
\includegraphics[scale=0.7]{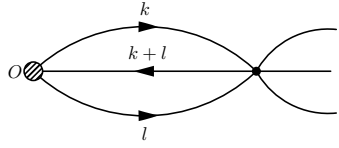}
\end{center}
\caption{\sl Two loop contribution of scalar sextet interaction
to anomalous dimension matrix of ${\cal O}$.}
\label{scalarsextet}
\end{figure}

We begin with the three-site scalar sextet contribution.
The Feynman integral
and numerical factors are all the same as the $M=N$ case
of Ref.\cite{Rey1}
 except $\lambda^2$ is now replaced by
$\lambda\bar\lambda$. The Feynman diagrams are depicted
in Fig.~\ref{scalarsextet}.
One may check that the diagram
involving the scalar sextet
potential has always one loop of scalar fundamental
and the other loop of scalar anti-fundamental.
Therefore, the contribution is of  mixed type and
 becomes
\be
H_{\rm B} = \lambda\bar\lambda
\sum_{\ell = 1}^{2L}  \Big[ {1 \over 2}
\mathbb{I}-  \mathbb{P}_{\ell,\ell+2} + {1 \over 2} \mathbb{P}_{\ell, \ell+2} \mathbb{K}_{\ell, \ell+1}
+ {1 \over 2} \mathbb{P}_{\ell, \ell+2} \mathbb{K}_{\ell+1, \ell+2}
-  {1 \over 2} \mathbb{K}_{\ell, \ell+1} \ \Big]\,.
\ee

\begin{figure}
\begin{center}
\includegraphics[scale=0.8]{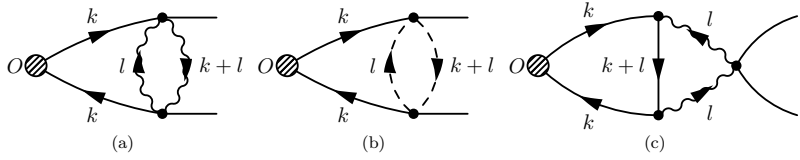}
\end{center}
\caption{\sl Two loop contribution of gauge
and fermion exchange interaction to anomalous dimension of ${\cal O}$.}
\label{feyn_05}
\end{figure}

Next we turn to the two-site gauge and fermion interactions.
As shown in Fig.~\ref{feyn_05}, there are three relevant non-vanishing
contributions. The first is the diamagnetic gauge diagram
contributing as a $\mathbb{I}$ type operator. There are one
 scalar loop and one gauge loop. One finds that the loop are always of the same
type, i.e. either $\lambda^2$ or $\bar{\lambda}^2$. For $M=N$ case the
contribution for each site was $-{\lambda^2\over 4} \mathbb{I}$.
Now one has an alternating contribution of
$ -{\lambda^2\over 4} \mathbb{I}$  and
$-{{\bar\lambda}^2\over 4} \mathbb{I}$ or
\be
H^{\rm gauge}_{\rm \,\, I}=
\bigl( {\lambda^2 +{\bar\lambda}^2}\bigr)
\sum_{\ell = 1}^{2L}  \Big[ \ - {1 \over 8}  \mathbb{I} \ \Big]\,.
\label{gaugeI}
\ee

On the other hand, the two site
fermion exchange contribution is always mixed type leading to the
$\mathbb{K}$ operator. There could be also mixed $\mathbb{I}$
type contribution in principle but they cancel among themselves
with the specific form of the Yukawa potential we have.
Therefore, the fermion two-site contribution becomes
\be
H_{\rm F}= {\lambda\bar\lambda} \sum_{\ell = 1}^{2L}  \mathbb{K}_{\ell, \ell+1}
\,.
\ee
The last diagram of Fig.~\ref{feyn_05} describes the two-site
gauge $\mathbb{K}$ type contribution. It is simple to check that
this contribution is of mixed type, whose expression reads
\be
H^{gauge}_{\rm \,\, K} = {\lambda\bar\lambda}
\sum_{\ell = 1}^{2L}  \Big[ \  - {1 \over 2}  \mathbb{K}_{\ell, \ell+1}
\ \Big].
\ee

We now turn to the contribution of the one-site interactions.
Adding up all the two-site interactions to the three-site interaction,
we see
that terms involving $\mathbb{K}$ operator cancel out one another.
So,
up to overall (volume-dependent) shift of the
ground state energy, the dilatation operator agrees
with the alternating spin chain Hamiltonian we derived.
 As we are dealing with
superconformal field theory, spectrum of
dilatation generator bears an absolute meaning.
 Therefore, to
 check the consistency with the supersymmetry,
we shall now compute terms arising from wave function renormalization
of $Y, Y^\dagger$. These are all the remaining contributions
to anomalous dimension of composite operator ${\cal O}$.

Wave function renormalization to $Y, Y^\dagger$ arises
from all three types of interactions. Even though there are
huge numbers of planar Feynman diagrams that could potentially
contribute to wave function renormalization,
many of them vanishes identically or cancel one another.

\begin{figure}
\begin{center}
\includegraphics[scale=0.8]{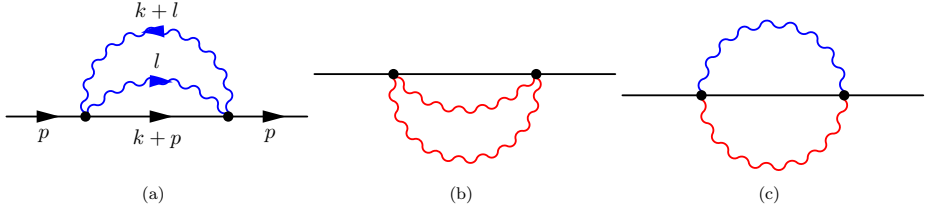}
\end{center}
\caption{\sl Two loop contribution of diamagnetic gauge
interactions to wave function renormalization of $Y, Y^\dagger$.
They contribute to $\mathbb{I}$ operator in the dilatation operator.}
\label{feyn_01}
\end{figure}

\begin{figure}
\begin{center}
\includegraphics[scale=0.5]{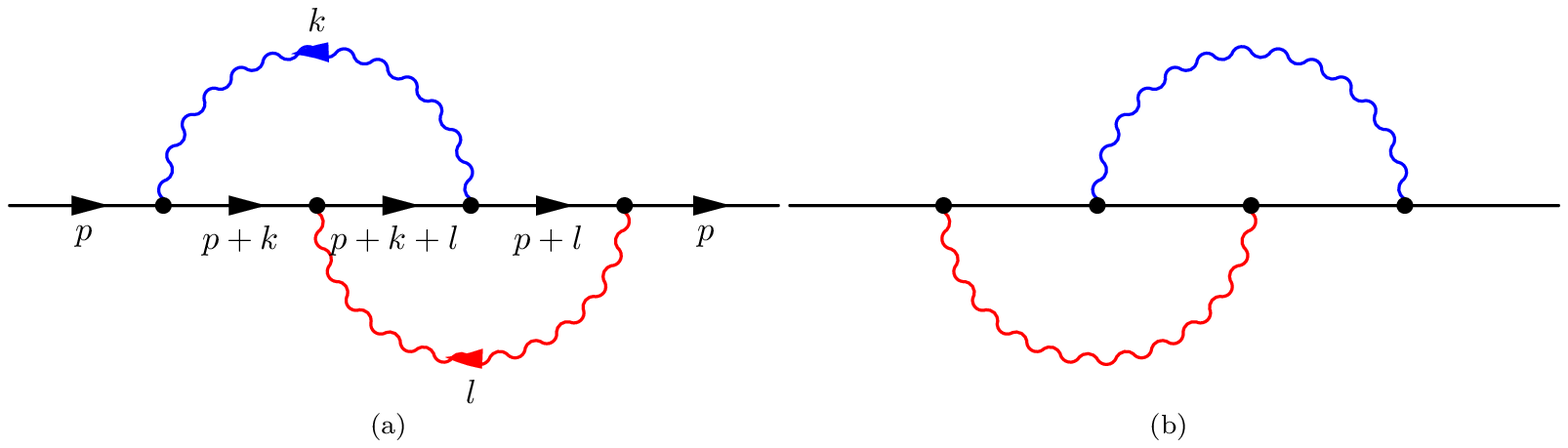}
\end{center}
\caption{\sl Two loop contribution of paramagnetic gauge
interactions to wave function renormalization of $Y, Y^\dagger$.
They contribute to $\mathbb{I}$ operator in the dilatation operator.}
\label{feyn_02}
\end{figure}

\begin{figure}
\begin{center}
\includegraphics[scale=0.5]{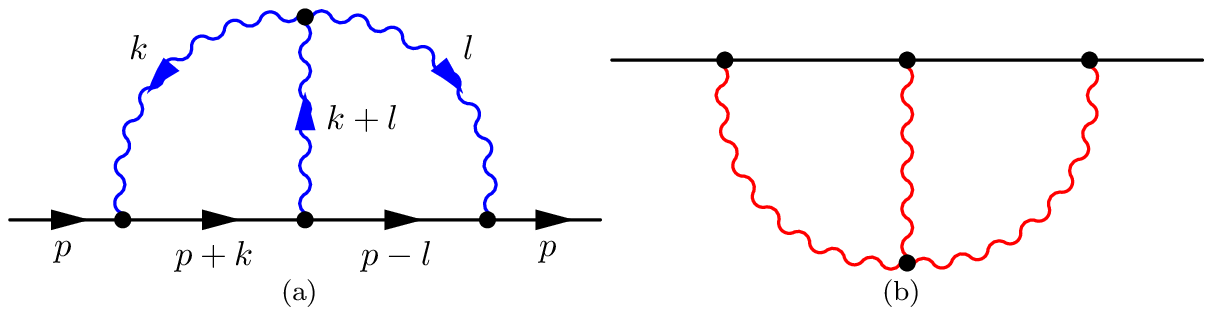}
\end{center}
\caption{\sl Two loop contribution of Chern-Simons interaction
to wave function renormalization of $Y, Y^\dagger$.
They contribute to $\mathbb{I}$ operators in the dilatation operator.}
\label{feyn_07}
\end{figure}

There are three types of
non-vanishing gauge contributions as shown
Figs.~\ref{feyn_01}-\ref{feyn_07}.
For the gauge diamagnetic contribution depicted in Fig.~\ref{feyn_01},
only the last one is of mixed type.  For the
$M=N$ case, each diagram
contributes  respectively by
$-{\lambda^2\over 24} {\mathbb{I}}$, $-{\lambda^2\over 24}
{\mathbb{I}}$
and ${\lambda^2\over 6} {\mathbb{I}}$
to the anomalous dimension for each site, which add up
to ${\lambda^2\over 12} {\mathbb{I}}$. For the present
case, their contribution is then
\be
H^{\rm dia}_{\rm \ Z} &=& \Big[ \ -{\lambda^2 \over 24}
 -{{\bar\lambda}^2 \over 24} +{\lambda\bar{\lambda} \over 6}
  \Big] \sum_{\ell=1}^{2L} \mathbb{I} \,.
\ee
The contributions of the gauge paramagnetic interaction in
Fig.~\ref{feyn_02} are obviously all mixed type. Hence, their
contribution becomes
\be
H^{\rm para}_{\rm \ Z} &=&
 {2 \lambda\bar{\lambda} \over 3}
  \sum_{\ell=1}^{2L} \mathbb{I} \,.
\ee
The contributions of the Chern-Simons interaction
in Fig.~\ref{feyn_07} are of types $\lambda^2$
or ${\bar\lambda}^2$. Its contribution now becomes
\be
H^{\rm cs}_{\rm \,\, Z} &=&
 \Big[ \ {\lambda^2 \over 6}
 +{{\bar\lambda}^2 \over 6}
  \Big]
  \sum_{\ell=1}^{2L} \mathbb{I} \,.
\ee

\begin{figure}
\begin{center}
\includegraphics[scale=0.8]{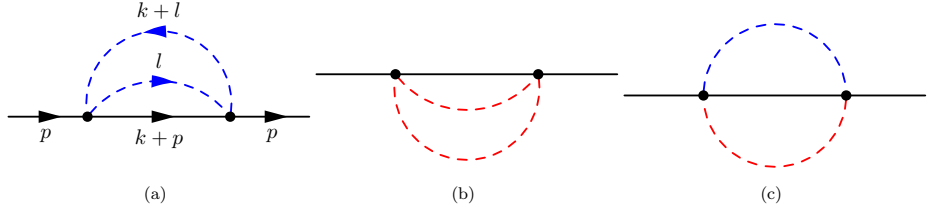}
\end{center}
\caption{\sl Two loop contribution of fermion pair interaction
 to wave function renormalization of $Y, Y^\dagger$.
They contribute to $\mathbb{I}$ operators in the dilatation operator.}
\label{feyn_03}
\end{figure}

\begin{figure}
\begin{center}
\includegraphics[scale=0.65]{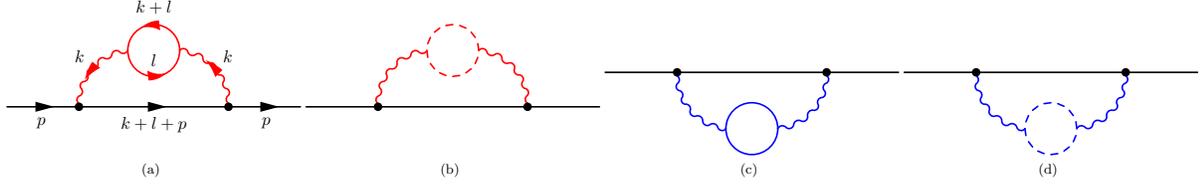}
\end{center}
\caption{\sl Two loop contribution of vacuum polarization to wave
function renormalization of $Y, Y^\dagger$.}
\label{feyn_04}
\end{figure}

The fermion pair interactions to the wave function renormalization
are depicted in Fig.~\ref{feyn_03} and they are all of mixed
type. Their contributions are
\be
H^{\rm yukawa}_{\rm \ Z} &=&
 \lambda\bar{\lambda} \Big[ \
  {4 \over 3} +1  \Big]
  \sum_{\ell=1}^{2L} \mathbb{I} \,.
\ee
Finally, we consider the two-loop contribution from
the vacuum polarization. Since the
Chern-Simons gauge loop and the corresponding ghost loop
contributions cancel with each other precisely, only the matter
loops have the non-vanishing contributions. The non-vanishing
two loop contributions of vacuum polarizations are all
mixed type, which are depicted in Fig.~\ref{feyn_04}.
Their contribution is
\be
H^{\rm vacuum}_{\rm \ Z} &=&
 -{8\lambda\bar{\lambda} \over 3}
  \sum_{\ell=1}^{2L} \mathbb{I} \,.
\ee
Summing up all these wave function renormalization to
$Y, Y^\dagger$, we find their contribution to
the anomalous dimension matrix as
\be
H_{\rm \,\, Z} &=& \Big[ \ {\lambda^2 \over 8}
 +{{\bar\lambda}^2 \over 8} +{\lambda\bar{\lambda} \over 2}
  \Big] \sum_{\ell=1}^{2L} \mathbb{I} \,.
\label{wavemixed}
\ee

One can see that the $\lambda^2$ and ${\bar\lambda}^2$ contributions
in (\ref{wavemixed}) cancel with those  of (\ref{gaugeI}).
Thus, one finds that only mixed type contributions remain.
Adding up all contributions,
\bea
H_{\rm total} &=& H_{\rm B} + H^{\rm gauge}_{\rm \,\, I}+H_{\rm F} +
H^{\rm gauge}_{\rm \,\, K} + H_{\rm Z} \eea
we get the result (\ref{twoloops}). As claimed, this is
 precisely the parity-symmetric
alternating spin chain Hamiltonian we obtained
from the mixed set of the relevant Yang-Baxter equations.

Finally, let us comment on the two loop wrapping interaction of
$L=1$ case as a checkpoint of internal consistency with ${\cal N}=6$
supersymmetry, extended to $M \ne N$.
The ${\bf 4}\otimes {\bf \bar{4}}$ representation is decomposed
irreducibly into the traceless
part, $\bf 15$, and the trace part, $\bf{1}$.
The multiplet $\bf 15$ is chiral primary operator,
so their conformal dimension ought to be protected by
supersymmetry. However there is no contribution of
three-site scalar interaction. Thus naively,
the protection of the above chiral primary operator is not
possible.  However, spectrum of the gauge invariant operator
of length $2L = 2$ will receive contributions from
wrapping diagrams already at leading
order, which we will identify.

From the above computations, the sum of  the two-site
and the one-site contributions
is
\be
H_2+ H_1 = \lambda{\bar\lambda} \Big[\  {1\over 2}
\mathbb{K}  + {1\over 2}
\mathbb{I}
\ \Big] \times 2 = \lambda{\bar\lambda} \Big[
\mathbb{K}  +
\mathbb{I}
\Big]
\,,
\ee
where the multiplication factor two comes from the
number of sites.

\begin{figure}
\begin{center}
\includegraphics[scale=0.6]{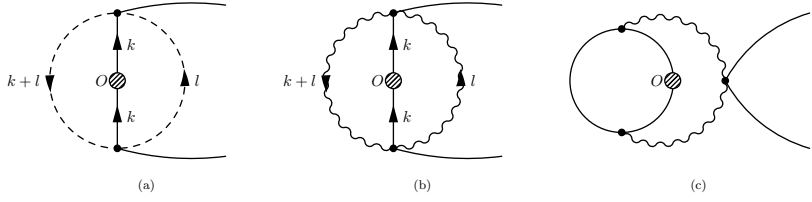}
\end{center}
\caption{\sl Two loop wrapping interaction contribution
to the shortest gauge invariant operators.
(a) fermion field wrapping,
(b) gauge field wrapping, (c) a new gauge triangle.}
\label{feyn_08}
\end{figure}

The wrapping contributions in Fig.\ref{feyn_08} are all mixed type.
The evaluation of the corresponding Feynman integrals are the same as
the case of $M=N$ except replacing $\lambda^2$ by $\lambda\bar\lambda$.
The results is
\be
H_{\rm wrap} = \lambda{\bar\lambda} \Big[\  \mathbb{I}
+2( \mathbb{K}  -
\mathbb{I}) - \mathbb{K}
\ \Big] = \lambda{\bar\lambda} \Big[
\mathbb{K}  -
\mathbb{I}
\Big]
\,,
\ee
Putting both the original and the wrapping diagram
contributions together, the full Hamiltonian of $2L=2$ operator is given by
\be
H_{2L=2} = 2 \,\lambda\bar\lambda \,\mathbb{K} \,.
\ee
Notice that the part proportional to $\mathbb{I}$ operator is
canceled between
the original and the wrapping interaction contributions.
One thus check that the chiral primary operators $\bf 15$ indeed
has a vanishing anomalous dimension since, by definition, it has
no trace part and is annihilated
by $\mathbb{K}$ operator. For the singlet ${\bf 1}$,
$|s\rangle= {1\over 2}|II\rangle$, the anomalous dimension is
\be
H |s\rangle= 8\,\,\lambda\bar\lambda\,\, |s\rangle\,.
\ee
So far, we computed the spectrum of the
shortest operators without a priori assumption of
supersymmetry. As a consistency check,
we now compare these spectra with their superpartners.
Recall that length $2 \ell$ operators with Dynkin labels
$(\ell-2m, m+n, \ell-2n)$ and length $2\ell-2$ operators
with Dynkin labels
$(\ell-2m, m+n-2, \ell-2n)$ are superpartners
each other. Here, we have the simplest situation:
the $L=1$ operator ${\bf 1}$ of Dynkin labels
$(0,0,0)$ is the superpartner of $L=2$ operator
${\bf 20}$ of Dynkin labels $(0,2,0)$.
Using the results of  Ref.~\cite{Minahan:2008hf},
the anomalous dimension of the latter can be found as
$8 \lambda\bar\lambda$, and matches perfectly with
our computation.

\section{Further Discussions}
The most salient feature of our results is that, though the Yang-Baxter
equations and hence the integrability structure permit it, the spin chain Hamiltonian
derived from the ABJ theory at two loops does not show parity symmetry breaking.
In this section, we elaborate further regarding this result and also provide
intuitive (albeit heuristic) argument for the reason why.

\begin{list}{$\bullet$}{}

\item {\bf Weak Coupling Limit: Closed Spin Chains} \hfill\break
The spin chain is, roughly speaking, weak coupling counterpart of the semiclassical string
propagating on AdS$_4 \times \mathbb{CP}_3$ with discrete $B_{\rm NS}$ holonomy. On the other hand, the supergravity dual background of the ABJ theory is given by
\be
\rmd s^2 &=& R^2_s  \left[ {1 \over 4} \rmd s^2
(\mbox{AdS}_4) + \rmd s^2 (\mathbb{CP}_3)  \right] \nonumber \\
e^{2 \phi} &=& {R_s^2 \over k^2} \nonumber \\
F_4 &=& {3 \over 8} k R_s^2 \widehat{\epsilon}_4 \nonumber \\
F_2 &=& k \rmd \omega = k \, J \nonumber \\
B_{\rm NS} &=& b \, J \, ,
\label{ABJbackground}
\ee
where $J$ is the K\"ahler two-form threading the $\mathbb{CP}^1$ inside $\mathbb{CP}^3$. Notice that the curvature radius is
\be
R^2_s = 2^{5\over 2} \pi \sqrt{\lambda} \, ,
\ee
is exactly the same as the background of ABJM theory, viz. the curvature radius remains {\sl unchanged} by turning on the $B_{\rm NS}$ holonomy. As such, the spectrum of light fields is unaffected by the discrete $B_{\rm NS}$ holonomy. This is consistent with ABJ's claim that the spectrum of all non-baryonic chiral primary operators is independent of $b$ but also goes beyond, asserting that {\sl all} string spectrum is
independent of the discrete holonomy.

Given that the spectrum of chiral primary operators is independent of $b$, it is not surprising that the spectrum of all single trace operators (\ref{STO}) is also independent of $b$ as well. Consider a closed, semiclassical string propagating in the background (\ref{ABJbackground}). The string is
macroscopic and propagates freely with the worldsheet topology of cylinder. This is the strong coupling counterpart of a single trace operator in the planar limit. Since the worldsheet has topology of cylinder, the integral over the pullback of the discrete $B_{\rm NS}$ holonomy would be zero.
ABJ argues further that, at strong `t Hooft coupling regime, all the U$(M)_k \times \overline{\rm U(N)}_{-k}$ theories with $M=N, N+1, \cdots, N+k$ are all similar to each other, since the only
difference is the discrete $B_{\rm NS}$ holonomy. Extrapolating this to the weak `t Hooft coupling regime, it then seems that all these theories are identical to all orders in the planar perturbation theory. Our result that the spin chain Hamiltonian of single trace operators is parity invariant fits to these ABJ arguments.

On the other hand, if the string trajectory wraps around $\mathbb{CP}_1$ over which the discrete $B_{\rm NS}$ holonomy is turned on, the integral will be nonzero. In fact, this leads to
the worldsheet instanton effect whose strength scales as $\exp (- \sqrt{\lambda})$. Transcribed
to the weak coupling limit, we conjecture that these worldsheet instanton effects may correspond to a class of unsuppressed fluctuations of the length of the single trace operators. These fluctuations are not generic ones, since they must be the counterpart of worldsheet topology of sphere. At present, though, it is unclear what precise nature of these fluctuations are.

Putting these considerations together, the $(M-N)$ dependent effect is completely suppressed at the strong `t Hooft coupling limit (modulo worldsheet instanton effect) and is most pronounced at the weak `t Hooft coupling limit, as reflected through the coupling parameter $\lambda \overline{\lambda} = \lambda^2 (1 + b/\lambda)$. Still, we found that the parity symmetry breaking effect, proportional to the sign of $(M-N)$, is invisible in the single trace operators.

\item {\bf Strong Coupling Limit: Giant Magnon} \hfill\break
Is the parity symmetry breaking visible at strong coupling limit,
$\lambda, \overline{\lambda} \rightarrow \infty$? Because of quantum
consistency, as discussed in Section 2, the coupling parameter $b$
is restricted to a discrete value ranging over $[0, 1]$. Therefore,
in the limit $\lambda, \overline{\lambda} \rightarrow \infty$, we
expect that parity symmetry breaking effect is completely suppressed
to the order ${\cal O}(1/\lambda, 1/\overline{\lambda})$.
Below, we confirm such expectation by demonstrating that the spectrum
of a giant magnon in the gravity dual of the ABJ theory is exactly the
same as that in the gravity dual of the ABJM theory.

We parametrize the $\mathbb{CP}_3$ metric as
\bea
\rmd s^2 &=& \rmd \xi^2 + {\sin^2 2\xi \over 4}
\left(
\rmd \psi + {\cos\theta_1\over 2} \rmd \phi_1 -
{\cos\theta_2\over 2} \rmd \phi_2
\right)^2
+ {1\over 4} \cos^2\xi
(\rmd \theta_1^2 +\sin^2\theta_1 \rmd \phi_1^2)\nonumber\\
&+& {1\over 4}\sin^2\xi (\rmd \theta_2^2 +\sin^2\theta_2 \rmd \phi_2^2) \, . 
\eea
The $B_{\rm NS}$ potential is
\be
&& B_{\rm NS}= -{b\over 2 
} \Bigl( {\sin 2\xi } \rmd \xi \wedge
\left(
 2\rmd \psi + {\cos\theta_1} \rmd \phi_1 -
{\cos\theta_2 } \rmd \phi_2
\right)\Bigr.
\nonumber\\
&& \ \ \ \  \Bigr.+ \cos^2\xi \sin\theta_1 \rmd \theta_1 \wedge \rmd \phi_1
+ \sin^2\xi \sin\theta_2 \rmd \theta_2 \wedge \rmd \phi_2 \Bigl) \, .
\ee

We work in the conformal gauge-fixing and choose the static gauge
$t =\tau$. We truncate the dynamics consistently on the first $\mathbb{S}^2$ by setting $\xi = 0$
and rename $\theta_1 = \theta, \phi_1 = \phi$. Bosonic part of the Type IIA superstring worldsheet action over $\mathbb{R}_t \times \mathbb{S}^2$ reads
\be
S=
\int \rmd \tau \int^r_{-r} \rmd \sigma \Big[{\pi\sqrt{2\lambda}\over 4\pi}
\Big( { (\partial z)^2\over 1-z^2}
+ (1-z^2)(\partial \phi)^2
\Big) + {b\over 4\pi
} (\dot{z}\phi' - {z'}\dot{\phi}  ) \Big]
\label{stringaction}
\ee
where $z= \cos\theta$.
In this set-up, the Virasoro constraints
\bea
&& {{\dot{z}}^2 + {z'}^2\over 1- z^2} + (1-z^2)({\dot{\phi}}^2 +
{\phi'}^2) = 1\,,\nonumber\\
&&
 {\dot{z}\,\, z'\over 1- z^2} + (1-z^2)\dot{\phi}\,\,
{\phi'} = 0
\eea
have to be imposed as well.
The energy density is uniform in the static  gauge
and the string energy is  proportional to
the spatial coordinate size:
\be
E= {\pi\sqrt{2\lambda}\over 2 \pi}\,\, 2r\,.
\ee

With an ansatz,
\be
z= z(\sigma - v \omega t)\,,\, \ \ \ \
\phi= \omega \tau + \varphi(\sigma - v \omega t)\,,
\ee
the equations of motion are reduced to
\bea
&& (z')^2 = {\omega^2\over (1- v^2\omega^2)^2} \left(z^2 - 1+ {1\over \omega^2}
\right) \left(1-v^2 -z^2\right)\nonumber\\
&& \varphi' = {v \omega^2\over (1- v^2\omega^2)} {z^2 - 1+ {1\over \omega^2}
\over 1-z^2}\,.
\label{eom}
\eea

The equation of motion is not affected by the $B_{\rm NS}$ field.
The worldsheet momentum $p$ is from the
$\int^{r}_{-r} (x_-)'$ with $x_- = t+ \phi$,
which equals to $\Delta\phi$. Hence it is independent of
$b$.
The expression for the
angular momentum is affected by
\be
J = \int^r_{-r}\left(
{\pi\sqrt{2\lambda}\over 4\pi}
(1-z^2)\dot\phi
-{b\over 4\pi
} z'
\right)
\ee
but, on the solution, its value does not change due to the
boundary condition of $z(-r)= z(r)$.
The general solution can found as~\cite{Zamaklar}
\be
z= {\sqrt{1-v^2} \over \omega \sqrt{\eta}} {\rm dn}\left( {\sigma- v \tau\over
\sqrt{\eta}\sqrt{1-v^2}}\,, \eta
\right)
\ee
where ${\rm dn}(\sigma, k^2)$ is the Jacobi elliptic function and
 we introduced the parameter $\eta$ by
\be
\eta = {1-\omega^2 v^2\over \omega^2 (1-v^2)}\,.
\ee
The range parameter $r$ is given by $\sqrt{1-v^2}\sqrt{\eta} K (\sqrt{\eta})$
where $K(x)$ is the complete elliptic integral. For simplicity, consider the infinite size limit
$\omega\rightarrow 1$~\footnote{It is trivial to extend the following analysis to a finite size
case.}.
The solution in this limit becomes
\be
z= \sqrt{1- v^2} {\rm sech}
\left({\sigma- v \tau \over \sqrt{1-v^2}}\right)\,.
\ee
with the worldsheet momentum given by  $p= 2 \cos^{-1} v$.
The spectrum
\be
E-J= \sqrt{2\lambda}\left|\sin {p \over 2}\right|\,,
\ee
remains unchanged, thus showing no $b$-dependence nor parity symmetry
breaking effect.

\item {\bf Weak Coupling Limit Revisited: Open Spin Chain} \hfill\break
Though effect of the discrete $B_{\rm NS}$ holonomy is invisible to closed
strings (up to the aforementioned worldsheet instanton effect), the holonomy
certainly affects spectrum of heavier string states such as D-branes that
wrap around $\mathbb{CP}_1$ over which the discrete $B_{\rm NS}$ holonomy is
turned on. These D-branes are giant gravitons and di-baryons and their excitation
is described by open strings attached to them.
Again, as for the closed string case, we see that the effect is suppressed in large `t Hooft coupling limit, while it could be pronounced in small `t Hooft coupling limit. From the
string worldsheet action (\ref{stringaction}), we expect that the boundary condition
gives rise to at most ${\cal O}(1/\lambda)$ effect.

Transcribed again to the weak `t Hooft coupling regime, a natural setting where
the parity symmetry breaking can be seen is the open spin chain attached to
giant gravitons or baryonic operators. The effect of $(M-N)$ should be reflected to
possible types boundary condition of the open spin chain. For example, since the
$B_{\rm NS}$ holonomy takes $(M-N)$ discrete values, we expect that there are
$(M-N)$ types of boundary conditions. For gauge group U$(M) \times \overline{{\rm U}(N)}$,
the baryonic operator $\epsilon_{a_1 \cdots a_M} \epsilon^{b_1 \cdots b_N} {Y^{I_1 a_1}}_{b_1} \cdots {Y^{I_N a_N}}_{b_N}$ is not a gauge singlet but transforms
as $(M-N)$-th antisymmetric product of fundamentals of the U$(M)$ gauge group. It is natural to expect that the $(M-N)$ types of open spin chain boundary conditions are associated with the multiplicity of these baryonic operator. For ${\cal N}=4$ super Yang-Mills theory,
such configuration of open spin chain was studied \cite{boundaries}. In fact,
boundary reflection matrices were determined for the tensor structure \cite{HM}
and for the dressing phases \cite{CC, ours}. We expect similar development can
be made in the ABJ theory with the new twist of the multiple boundary conditions. We are currently investigating this and will report the results elsewhere.
\end{list}

Finally, since the spin chain Hamiltonian of the ABJ theory takes the same form as
the ABJM theory, diagonalization of the transfer matrices proceeds the same manner.
Thus, the Bethe ansatz equations of SO(6) sector \cite{Minahan:2008hf, Rey1} and of full
OSp$(6 \vert 4, \mathbb{R})$ \cite{Minahan:2008hf} will have exactly the same form
except that $\lambda^2$ of the ABJM theory counterpart is now replaced by $\lambda \overline{\lambda}$.

\section*{Acknowledgement}
We are grateful to Ofer Aharony for many useful correspondences and
to Matthias Staudacher for many illuminating discussions. We also thank
Anamaria Fonts and Stefan Theisen for discussions on several issues
related to discrete torsion.
This work was supported in part by R01-2008-000-10656-0 (DSB),
SRC-CQUeST-R11-2005-021 (DSB,SJR), KRF-2005-084-C00003 (SJR),
EU FP6 Marie Curie Research \& Training Networks MRTN-CT-2004-512194
and HPRN-CT-2006-035863 through MOST/KICOS (SJR), and F.W.
Bessel Award of Alexander von Humboldt Foundation (SJR).

\appendix

\section{${\cal N}=6$ ${\rm U(M)}\times \overline{\rm U(N)}$
 Super Chern-Simons
Theory}
$\bullet$ Gauge and global symmetries:
\bea
&& \mbox{gauge symmetry}: \quad \mbox{U(M)} \otimes \overline{\mbox{U(N)}}
\nonumber \\
&& \mbox{global symmetry}: \quad \mbox{SU(4)}
\eea
We denote trace over U(M) and $\overline{\rm U(N)}$
as Tr and $\overline{\rm Tr}$, respectively.

$\bullet$ On-shell fields are gauge fields,
complexified Hermitian scalars and Majorana spinors ($I=1,2,3,4$):
\bea
&& A_m: \quad \mbox{Adj}\,\,\, (\mbox{U(M)}); \hskip2cm \overline{A}_m : \quad
\mbox{Adj}\,\,\, \overline{\mbox{U(N)}} \nonumber \\
&& Y^I = (X^1 + i X^5, X^2 + i X^6, X^3 - i X^7, X^4 - i X^8):
\qquad ({\bf M}, \overline{\bf N}; {\bf 4}) \nonumber \\
&& Y^\dagger_I = (X^1 - i X^5, X^2 - i X^6, X^3 + i X^7, X^4 + i X^8):
\hskip0.8cm ( \overline{\bf M}, {\bf N}; \overline{\bf 4})
\nonumber \\
&& \Psi_I = (\psi^2 + i \chi^2, -
\psi^1 - i \chi^1, \psi_4  - i \chi_4 , - \psi_3 + i \chi_3 ) : \hskip0.4cm
({\bf M}, \overline{\bf N};
\overline{\bf 4}) \nonumber \\
&& \Psi^{\dagger I} = (\psi_2 - i \chi_2, -
\psi_1 + i \chi^1, \psi^4 + i \chi^4, - \psi^3 - i \chi^3): \hskip0.3cm
(\overline{\bf M}, {\bf N}; {\bf 4})
\eea
$\bullet$ action:
\bea
I &=&   \int_{\mathbb{R}^{1,2}}
\Big[ \, {k \over 4\pi} \epsilon^{mnp} \mbox{Tr} \left(A_m \partial_n A_p
+{2 i \over 3} A_m A_n A_p \right)
- {k \over 4\pi} \epsilon^{mnp} \overline{\mbox{Tr}}
\left(  \overline{A}_m \partial_n \overline{A}_p +{2 i \over 3} \overline{A}_m
\overline{A}_n \overline{A}_p \right) \nonumber \\
&& \hskip1cm +{1 \over 2} \overline{\mbox{Tr}}
\left( -(D_m Y)^\dagger_I D^m Y^I  + i \Psi^{\dagger I} D
\hskip-0.22cm / \Psi_I  \right) + {1 \over 2} \mbox{Tr}
\left(- D_m Y^I (D^m Y)^\dagger_I  +
 i \Psi_I D \hskip-0.22cm / \Psi^{\dagger I}  \right)
\nonumber \\
&&
\hskip1cm - V_{\rm F} - V_{\rm B} \, \Big]
\eea
Here, covariant derivatives
are defined as
\bea
D_m Y^I = \partial_m Y^I + i A_m Y^I - i Y^I \overline{A}_m \, , \quad D_m Y^\dagger_I = \partial_m Y^\dagger_I + i \overline{A}_m Y^\dagger_I - i Y^\dagger_I A_m
\eea
and similarly for fermions $\Psi_I, \Psi^{\dagger I}$. Potential terms are
\bea
V_{\rm F} &=& {2 \pi i \over k} \overline{\mbox{Tr}}
 \Big[ Y^\dagger_I Y^I \Psi^{\dagger J} \Psi_J
- 2 Y^\dagger_I Y^J  \Psi^{\dagger I} \Psi_J  +
\epsilon^{IJKL} Y^\dagger_I \Psi_J Y^\dagger_K \Psi_L]
\nonumber \\
&-& {2 \pi i \over k} \mbox{Tr} [Y^I Y^\dagger_I \Psi_J  \Psi^{\dagger J}
- 2 Y^I Y^\dagger_J \Psi_I \Psi^{\dagger J} + \epsilon_{IJKL}
 Y^I \Psi^{\dagger J} Y^K \Psi^{\dagger L} \Big]
\eea
and
\bea
V_{\rm B} &=& - {1 \over 3} \left({2 \pi \over k}\right)^2
\overline{\mbox{Tr}} \Big[ \, Y^\dagger_I Y^J Y^\dagger_J Y^K Y^\dagger_K Y^I
+ Y^\dagger_I Y^I Y^\dagger_J Y^J Y^\dagger_K Y^K \nonumber \\
&& \hskip1.8cm + 4 Y^\dagger_I Y^J Y^\dagger_K
Y^I Y^\dagger_J Y^K -
6 Y^\dagger_I Y^I Y^\dagger_J Y^K Y^\dagger_K Y^J \, \Big]
\eea
At quantum level, since the Chern-Simons term shifts by an integer multiple of
$8 \pi^2$, $k$ should be integrally quantized.

\end{document}